\pdfoutput=1
\documentclass[%
reprint,
superscriptaddress,
twocolumn,
bibnotes,
amsmath,amssymb,
]{revtex4-2}
\usepackage{footmisc}

\usepackage{graphicx}
\usepackage{dcolumn}
\usepackage{bm}
\usepackage{titlesec}
\titleformat{\section}{\raggedright\fontsize{12.5}{25}\bfseries}{\arabic{section}.}{1em}{}

\begin{document}

\title{\fontsize{15}{19}\selectfont Investigation of the Non-equilibrium State of Strongly Correlated  Materials by Complementary Ultrafast Spectroscopy Techniques}

\author{H. Hedayat}
\affiliation{IFN-CNR, Dipartimento di Fisica, Politecnico di Milano, 20133 Milan, Italy}
\affiliation{Dipartimento di Fisica, Politecnico di Milano, 20133 Milan, Italy}
\author{C. J. Sayers}
\affiliation{Dipartimento di Fisica, Politecnico di Milano, 20133 Milan, Italy}
\affiliation{Department of Physics and Centre for Photonics and Photonic Materials, University of Bath, BA2 7AY Bath, UK}

\author{A. Ceraso}
\affiliation{IFN-CNR, Dipartimento di Fisica, Politecnico di Milano, 20133 Milan, Italy}
\affiliation{Dipartimento di Fisica, Politecnico di Milano, 20133 Milan, Italy}

\author{J. van Wezel}
\affiliation{Institute for Theoretical Physics, Institute of Physics, University of Amsterdam, 1090 GL Amsterdam, The Netherlands}
\author{S. R. Clark}
\affiliation{H. H. Wills Physics Laboratory, University of Bristol, BS8 1TL Bristol, UK}
\author{C. Dallera}
\affiliation{Dipartimento di Fisica, Politecnico di Milano, 20133 Milan, Italy}
\author{G. Cerullo}
\affiliation{Dipartimento di Fisica, Politecnico di Milano, 20133 Milan, Italy}

\author{E. Da Como}
\affiliation{Department of Physics and Centre for Photonics and Photonic Materials, University of Bath, BA2 7AY Bath, UK}

\author{E. Carpene}
\email{ettore.carpene@polimi.it}
\affiliation{IFN-CNR, Dipartimento di Fisica, Politecnico di Milano, 20133 Milan, Italy}

\begin{abstract}
 Photoinduced non-thermal phase transitions are new paradigms of exotic non-equilibrium physics of strongly correlated materials. An ultrashort optical pulse can drive the system to a new order through complex microscopic interactions that do not occur in the equilibrium state. Ultrafast spectroscopies are unique tools to reveal the underlying mechanisms of such transitions which lead to transient phases of matter. Yet, their individual specificities often do not provide an exhaustive picture of the physical problem.  One effective solution to enhance their performance is the integration of different ultrafast techniques. This provides an opportunity to simultaneously probe physical phenomena from different perspectives whilst maintaining the same experimental conditions. In this context, we performed complementary experiments by combining time-resolved reflectivity and time and angle-resolved photoemission spectroscopy. We demonstrated the advantage of this combined approach by investigating the complex charge density wave (CDW) phase in 1$\it{T}$-TiSe$_{2}$. Specifically, we show the key role of lattice degrees of freedom to establish and stabilize the CDW in this material.
\end{abstract}

\maketitle
\section{Introduction}
\label{sec:intro}  

Ultrafast spectroscopies are powerful tools to unveil the underlying physics that dictates the properties of  strongly correlated electron
systems~\cite{shah2013ultrafast}. In a conventional pump-probe scheme, an ultrashort optical pulse (pump) triggers a cascade of interactions among internal degrees of freedom (electronic, lattice, orbital and spin) and a second time-delayed pulse (probe) provides information about their dynamics. In most cases, the information is collected from the reflected (or transmitted) photons or photoemitted electrons. Accordingly, the techniques can be classified as time-resolved optical spectroscopy ~\cite{dal2015snapshots,muller2009spin} or photoemission spectroscopy ~\cite{smallwood2012tracking,saule2019high}, respectively. Each of these broad categories includes numerous variants to provide more details on the out-of-equilibrium state of matter. Because of the great complexity of the underlying microscopic interactions, no single experimental approach can provide a comprehensive picture of all subsystem dynamics after perturbation. Revealing the complex electron and quasiparticle (QP) interactions ~\cite{sentef2013examining,hoyer2005many}, the extreme competition of phases due to the interplay between order parameters~\cite{demsar1999superconducting,schmitt2008transient}, transient photo-induced creation of new phases or states ~\cite{fausti2011light,stojchevska2014ultrafast,wang2013observation} and many other exotic effects is beyond the capability of any single technique alone. Thus, complementary ultrafast techniques are necessary for gaining a clearer understanding of the excited states in correlated materials. Although newer studies normally take into account the earlier results using complementary techniques to build a consistent physical picture, the comparison of individual experiments is not always straightforward. Because of the complexity of experimental methods, even a slight difference in the conditions makes the comparison of two separate outcomes challenging. Due to the large number of experimental parameters, such as the sample temperature, pump wavelength, polarization, sample inhomogeneity, defects, doping, etc., the outcomes of different experiments are often difficult to compare. To improve reliability and make comparisons straightforward, one should reduce the diverging aspects of different experiments as much as possible. With this in mind, the integration of complementary techniques is an innovative approach to propel the advancement of ultrafast spectroscopies and to widen their capabilities  ~\cite{gerber2017femtosecond,zong2019evidence,hedayat2019excitonic}. The aim is to use various probes, but under identical perturbation and environmental conditions in order to disclose different aspects of the same phenomena. Here, we show that the integration of time-resolved reflectivity (TRR) and time and angle-resolved photoemission spectroscopy (TR-ARPES) allows to shed new light on the exotic physics of 1$\it{T}$-TiSe$_{2}$. Our study demonstrates the fundamental role of excitons and phonons in the charge density wave (CDW) phase of 1$\it{T}$-TiSe$_{2}$.  \\
TRR records photo-induced changes of the sample reflectivity ($\Delta R$) resulting from the variation of its dielectric function $ \epsilon (\omega )={\epsilon}' (\omega )+i{\epsilon}''(\omega )$~\cite{vivsvnovsky1984magnetooptical,carpene2015ultrafast}, where $\omega$ is the frequency of the probe. $R(\omega)$ is a function of $ \epsilon (\omega )$, experimental geometry and beam polarization (the two last terms are kept constant during experiments). For example, for a non-magnetic sample with cubic symmetry  at normal incidence, $R(\omega) = |(1- \sqrt{{\epsilon(\omega )}}) / (1 + \sqrt{{\epsilon(\omega )}})|^{2}$~\cite{boschini2015flexible}. Any pump-induced electronic modification alters the dielectric function, and thus the TRR signal. Assuming a linear response and using a simple first-order perturbation approach, the pump-induced modifications of carrier density ($n_{e}$), the electronic temperature ($T_{e}$), lattice coordinate ($q$) and any other relevant parameter can be made explicit in the transient reflectivity signal $\Delta R(t)$, through the dielectric function, as follows~\cite{placzek1959rayleigh,zeiger1992theory}

\begin{multline}
\Delta R(t)=\frac{\partial R}{\partial \epsilon}\frac{\partial \epsilon}{\partial n_{e}}\Delta n_{e} (t)+\frac{\partial R}{\partial \epsilon}\frac{\partial \epsilon}{\partial T_{e}}\Delta T_{e}(t)\\
+\frac{\partial R}{\partial \epsilon}\frac{\partial \epsilon}{\partial q}\Delta q(t).\label{trr1}
\end{multline} 

A quantitative estimate of $\Delta R(t)$ requires the knowledge of $\epsilon(n_{e}$, $T_{e}$,  $q$) which is not an easy task and is beyond the purpose of this work. However, Eq~\ref{trr1} clearly shows how the TRR signal can provide substantial information on electron dynamics~\cite{sun1993femtosecond,hirori2003electron}, electron-phonon coupling strength~\cite{gadermaier2010electron,mansart2010ultrafast} and lattice motions~\cite{yusupov2010coherent}. The last term in particular shows that using TRR one can detect the non-equilibrium coherent lattice modes~\cite{sayers2020coherent}, their excitation mechanisms, lifetime, amplitude~\cite{alfano1971optical} or even quantitatively determine atomic displacements~\cite{decamp2001dynamics} after photoexcitation. Unveiling the role of the lattice in properties of quantum materials has been the subject of several investigations due to its great potential, e.g. for the coherent control of non-thermal phase transitions ~\cite{cavalleri2001femtosecond,rousse2001non}. For this purpose, TRR is an effective technique due to its excellent signal-to-noise ratio combined with femtosecond temporal resolution. After the pump, excited carriers and QPs relax by different decay mechanisms and emitting coherent phonons~\cite{schmitt2008transient}. We consider that the source of displacive phonon excitation is the decay of photoexcited carriers and QPs, by assuming a linear dependence of the quasi-equilibrium lattice coordinate $q_{0}$, as $q_{0}(t)=\kappa\,.\,n_{e}$ in Eq~\ref{trr1}. Thus, the coherent structural vibrations part of TRR can be separated from incoherent electronic and QP part as~\cite{zeiger1992theory,mertelj2013incoherent} 
\begin{multline}
\Delta R (t)=[C_{1}\,\mathrm{exp}(-t/\tau_{e} )+C_{2}]\\
+\sum_{i} C_{3,i}\,\mathrm{exp}(-t/\tau_{d,i})[\mathrm{cos}(t/T_{i})-\beta _{i}\mathrm{sin} (t/T _{i})],\label{trr2}
\end{multline} 
where $\beta_{i} =(1/\tau_{e}-1/\tau_{d,i})T _{i}$ and pump and probe pulses are sufficiently short compared to the periods of oscillations. The first term in Eq~\ref{trr2} corresponds to the incoherent electron and QP relaxation with the time constant $\tau_{e}$. The sum in the second term is over all excited normal modes of coherent phonons with damping constants $\tau_{d,i}$ and periods $T _{i}$. $C_{1}$ and $C_{3,i}$ correspond to the QPs and phonons amplitudes, respectively, and $C_{2}$ is a constant background of lattice contribution for long delays. If the QPs decay times $\tau_{e}$ is not too short, $\beta_{i}$ becomes negligible and the oscillation can be described by a simple $\mathrm{cos}(t/T_{i})$ phase.
\begin{figure*}
\centering
\includegraphics[scale=0.57]{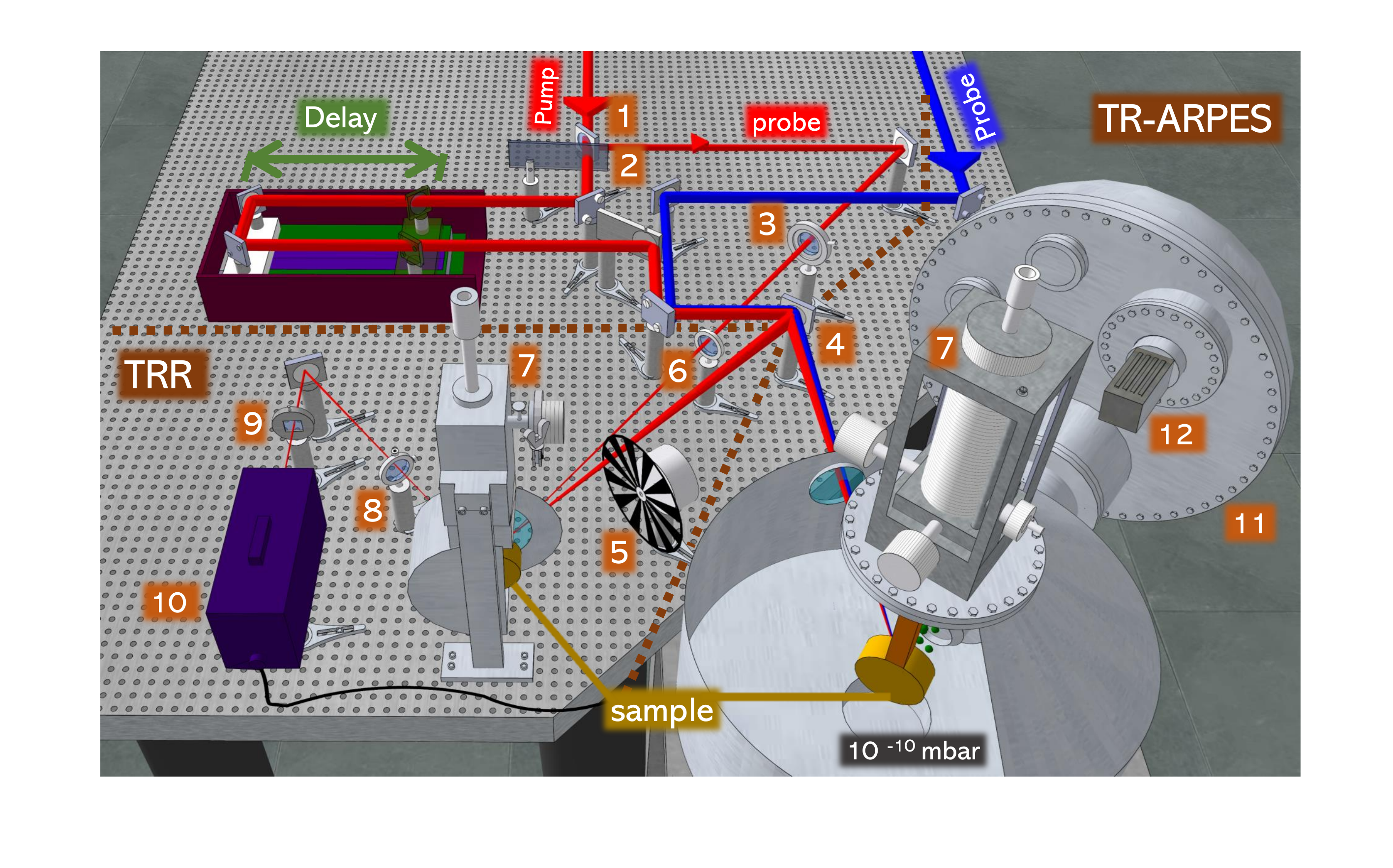}
\caption{Sketch of the system to perform combined TRR and TR-ARPES experiments. The optical layout to generate the pump (red beam) via the NOPA stage and probe (blue beam) via SFG is explained elsewhere~\cite{boschini2014innovative}. The common part is the excitation system which consists of tunable pump pulses and a motorized delay stage. The remaining two sections separated by dashed lines show the components for TRR or TR-ARPES experiments. Each experiment is carried out separately and one can switch between them by rotating and replacing the spherical mirror (number 4). The other numbered components correspond to;  (1)~beam splitter (2)~optical attenuator (3)~half-wave plate (4)~spherical mirror (5)~chopper (6)~focusing lens (7)~cryostat (8)~recollimating lens (9)~Glan polarizer (10)~photodiode (11)~hemispherical analyzer (12)~CCD camera. }
\label{Fig1}
\end{figure*}

Although TRR gives valuable information on non-equilibrium behavior of the electron and lattice systems, it does not provide a complete understanding of the interactions since it crucially lacks crystal momentum resolution. Momentum-sensitive techniques can clarify the microscopic mechanisms by resolving specific electron states and their interactions with other degrees of freedom. Indeed, TR-ARPES has been effectively exploited to study the evolution of order parameters in different complex systems ~\cite{perfetti2007ultrafast,gierz2013snapshots,hedayat2018surface}. In contrast to TRR, the technique of TR-ARPES probes the electronic band structure in the reciprocal space enabling one to focus on a specific region of interest e.g. across an energy gap. The total photoemission intensity is given by~\cite{damascelli2003angle}
\begin{align}
I(\omega,\mathbf{k},t)= M^{2}(\mathbf{\omega,k})[A(\omega,\mathbf{k},t)f(\omega,t)]+B(\omega,\mathbf{k},t),\label{pes1}
\end{align}
where $\it{M}(\mathbf{\omega,k})$ is the matrix element between the initial and final states determined by the probe photons, $\it{A}(\omega,\mathbf{k},t)$ is the single-particle spectral function, $\it{f}(\omega,t)$ is the Fermi Dirac distribution, and $\it{B}(\omega,\mathbf{k},t)$ is the background due to inelastic scatterings of photoelectrons. Therefore, TR-ARPES provides the time evolution of $A(\omega,\mathbf{k})$ which is a function of the self-energy, $\Sigma (\omega ,\mathbf{k})=\Sigma{}' (\omega ,\mathbf{k})+i\Sigma{}'' (\omega ,\mathbf{k}) $ and bare dispersion, and includes all information on the electronic band dispersion and (quasi) particle interactions. It also reveals the electron-phonon coupling by detecting the effect of lattice vibrations on the electronic states via deformation potential~\cite{khan1984deformation}. The energy and momentum sensitivity allows to uncover the coherent lattice vibrations which are coupled to specific electron states or the order parameter of a phase, i.e. selectively coupled phonons (SCPs)~\cite{sayers2020coherent,shi2019ultrafast,schmitt2011ultrafast,sobota2014distinguishing}. Thus, $A(\omega,\mathbf{k},t)$ disentangles the specific electron dynamics or coherent phonons which play a direct role in the establishment of order, such as superconductivity or CDWs~\cite{cortes2011momentum,hellmann2012time}. Although extracting the dynamics of electrons, energy gaps, QPs, SCPs and phonons is potentially feasible by the TR-ARPES technique, in practice, depending on the features of the apparatus and the material under investigation, it is very challenging to provide information on all those ultrafast phenomena together. TR-ARPES systems with probe photon energy of about 6 eV can only measure a limited portion of the Brillouin Zone (BZ). In TR-ARPES setups with higher probe photon energy (20-30 eV), granting access to the entire surface BZ, attaining an excellent combination of time and energy resolution with a high photon flux of the probe is extremely challenging~\cite{liu2017femtosecond}. Moreover, to obtain time and momentum resolution along the direction normal to the surface, one should acquire data for several probe photon energies. In general, to study highly complex materials, current TR-ARPES systems have limitations that should be considered. For example, resolving minor energy oscillations caused by weak phonons modulating the excited bands in different points of the BZ, if possible, often requires extensive acquisition times. This leads to other complications such as gradual surface contamination reducing the quality of obtained data. On the other hand, in TRR, an excellent signal-to-noise ratio and high temporal resolution are more easily achievable. TRR then allows investigation of a weak coherent response of excited lattice coupled to electronic states anywhere (although unknown) in the BZ.

To benefit from both techniques, we developed an experimental setup that provides the opportunity to perform TRR and TR-ARPES with the same pump pulses. Therefore, the complementary experiments probe similar photoexcitation and relaxation dynamics with minimum divergence in experimental conditions. One can then extract and compare the time evolution of $\epsilon (\omega)$ and $A(\omega,\mathbf{k})$ providing a richer information on the interactions between different subsystems. In order to demonstrate the capabilities of such a combined system, we performed experiments on 1$\it{T}$-TiSe$_{2}$, due to the complex nature of its CDW phase (see Sec~\ref{sec3} for more details). We show that the combination of TRR and TR-ARPES provides a new benchmark to investigate the complex mechanisms in strongly correlated materials.
\section{Combined ultrafast techniques}
\begin{figure*}
\centering
\includegraphics[scale=1.3]{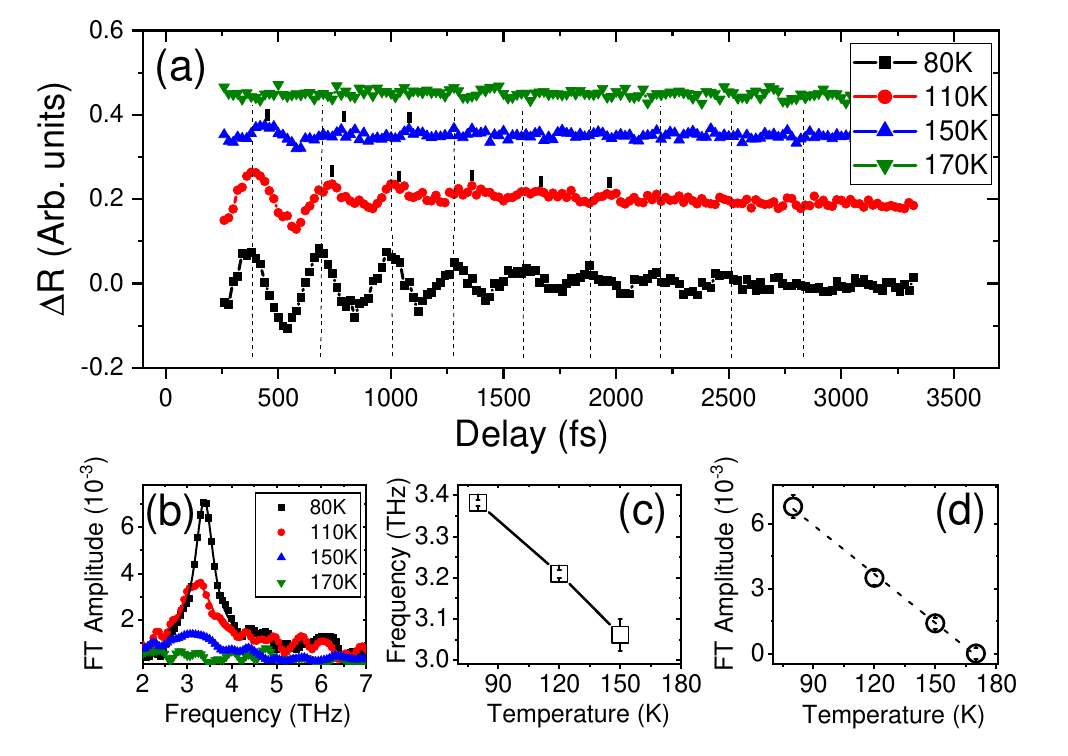}
\caption{(a) The changes in the oscillatory part of reflectively signal $\Delta$R of 1$\it{T}$-TiSe$_{2}$ as a function of pump-probe delay for different temperatures, $80$, $110$, $150$ and $170$K. The excitation density is  $2.77\times 10^{18}$~Cm$^{-3}$. Dashed lines show the oscillation peaks at $80$K, also the peaks of higher temperatures are marked. (b) Fourier transform (FT) of $\Delta R$ dynamics at different temperatures presented in Fig~\ref{Fig4}. (c) Oscillation frequency as a function of temperature, obtained by Gaussian fitting of FTs in panel (b). (d) Oscillation amplitude as a function of temperature. The dashed line shows the linear relation between temperature and amplitude. Error bars are scaled with the fitting precision.}
\label{Fig2}
\end{figure*}
\begin{figure*}
\centering
\includegraphics[scale=1.3]{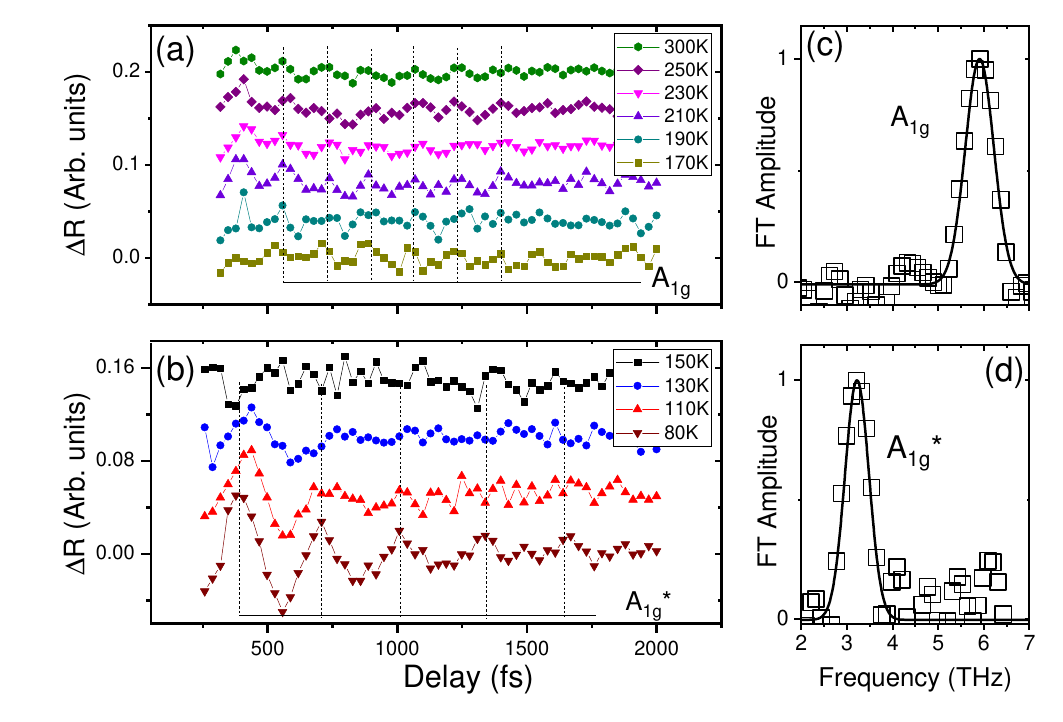}
\caption{TRR measurements with $6.45\times 10^{18}$~cm$^{-3}$ pump photoexcitation density and for a wide range of temperatures crossing $\textit{T}_{\mathrm{CDW}}$. Two distinct TRR response (a) and (b) is observed for temperatures below and above $\textit{T}_{\mathrm{CDW}}$. (c) and (d) are the normalized amplitude of Fourier transforms (FT) of the 80~K and 210~K oscillatory components. $T<\textit{T}_{\mathrm{CDW}}$ corresponds to the $\textit{A}^{*}_{\textrm{1g}}$ CDW phonon mode (3.19 THz) whilst for $T>\textit{T}_{\mathrm{CDW}}$, the $\textit{A}_{\textrm{1g}}$  normal phase mode (5.95 THz).}
\label{Fig3}
\end{figure*}

\begin{figure*}
\centering
\includegraphics[scale=1.3]{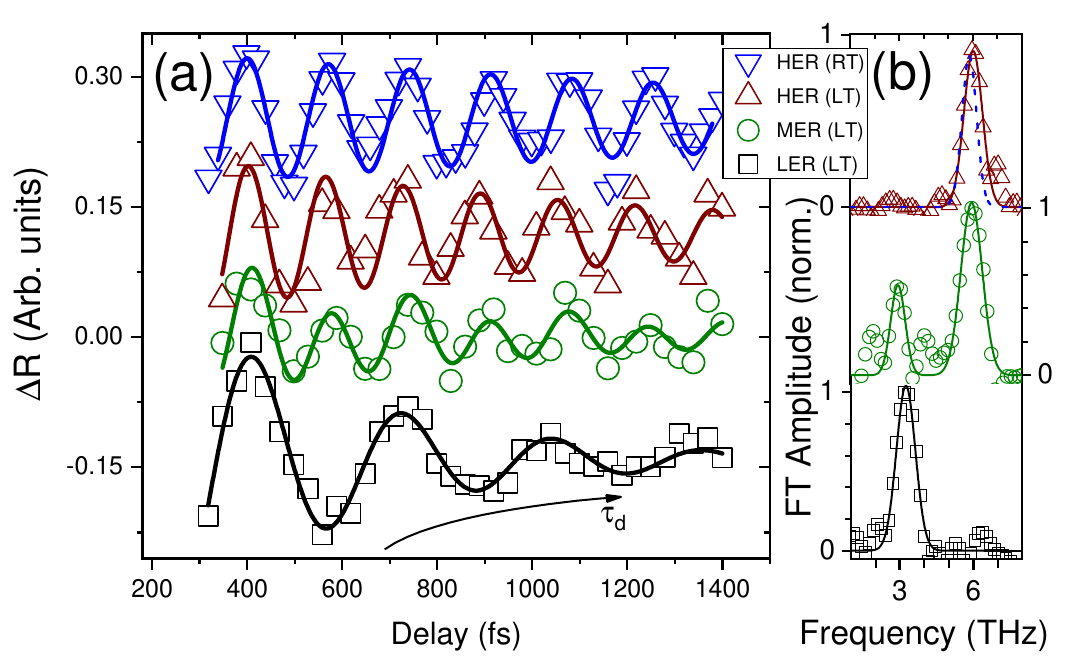}
\caption{(a) The oscillatory part of the TRR response for low $1.01\times 10^{19}$~cm$^{-3}$, medium $7.83\times 10^{19}$~cm$^{-3}$ and high $1.21\times 10^{20}$~cm$^{-3}$ excitation densities at 80~K (low temperature, LT) and excitation density of $2.43\times 10^{20}$~cm$^{-3}$ at 300~K (room temperature, RT) for comparison. The solid lines are fits to extract the damping time and oscillation frequency.  Panel (b) shows the FTs of the oscillations of TRR signal for low, medium and high excitation regimes (LER, MER and HER). MER is the crossover between the dominant $\textit{A}^{*}_{\textrm{1g}}$ and $\textit{A}_{\textrm{1g}}$ phonon oscillations at LER and HER, therefore the phonons are switched at $n_{\mathrm{th}}\approx7-8\times 10^{19}$~cm$^{-3}$. Gaussian fits of the FTs give frequencies of 3.95 THz and 5.95 THz for two excitation regimes corresponding to $\textit{A}^{*}_{\textrm{1g}}$ and $\textit{A}_{\textrm{1g}}$ phonon modes. The dashed blue line shows the FT of the TRR response at 300~K for comparison.}
\label{Fig4}
\end{figure*}
To perform TR-ARPES and TRR experiments, we used a commercial Yb-based system (Pharos, Light Conversion) generating 300 fs pulses with wavelength $\lambda = 1030$~nm at 80~kHz repetition rate. Ultrashort visible pulses of 30 fs are generated by a non-collinear optical parametric amplifier (NOPA) followed by pulse compression. The NOPA central wavelength is tunable in the range from 620 to 720~nm. Figure~\ref{Fig1} illustrates the experimental setup. The NOPA pulses are used as a pump for both TR-ARPES and TRR setups (thick red beam). A small fraction of the pump is divided by a beam splitter (no.~1 in Fig~\ref{Fig1}) and serves as the TRR probe (thin red beam).  A deep ultraviolet probe for TR-ARPES with a wavelength of about 206 nm and a duration of 65 fs (blue beam) is generated via second-harmonic generation of the NOPA output followed by sum-frequency generation (SFG) with the second harmonic of the 1030-nm beam, using a $\beta$-barium-borate (BBO) crystal. Further details on pump and probe pulse generation in TR-ARPES can be found in Ref.~\cite{boschini2014innovative}. In both experiments, a mechanical translation stage controls the delay between pump and probe pulses. Experiments are performed separately and can be switched by rotating and replacing the curved mirror (no.~4). The sample temperature in both experiments is controlled using cryostats over the range (10 - 500)~K (no.~7). The polarization of beams can be controlled by waveplates. In TRR, cross-polarized beams are typically used to reduce pump scattering artifacts, this is achieved by rotating the probe polarization (no. 3). The electrical signals from the photodiode detector (no.~10) are fed into a lock-in amplifier triggered at the frequency of a chopper (no.~5) on the path of the pump. TR-ARPES measurements are carried out in a $\mu$-metal shielded chamber equipped with a home built time of flight (ToF) analyzer in the previous version of the setup~\cite{carpene2009versatile}, which has been recently upgraded with (no.~11) a hemispherical energy analyzer (Phoibos 100, Specs). The TR-ARPES data shown here refer to the ToF energy analyzer. The overall time and energy resolution of the system are about 85 fs and 45 meV, respectively. Before TR-ARPES measurements, the samples are cleaved \textit{in-situ} at a pressure better than $5\times 10^{-10}$~mbar and surface quality and crystallographic orientation are checked by low energy electron diffraction (LEED). 

\section{Case-study: CDW in 1$\it{T}$-TiSe$_{2}$}\label{sec3}
Using the above described complementary ultrafast techniques, we investigated the non-thermal CDW phase transition in 1$\it{T}$-TiSe$_{2}$. The complex origin of CDWs in 1$\it{T}$-TiSe$_{2}$ has been at the center of a long-standing debate~\cite{zunger1978band,di1976electronic,hughes1977structural}. The formation of CDWs in this material is accompanied by the opening of a gap in the electronic states and a periodic lattice distortion (PLD).  Below 202~K, the two phenomena progress simultaneously, and therefore it is a major challenge to understand if either process is the primary driving force and the other is simply a consequence.  In the electronic structure, the VB shifts to lower energies, and a gap of about 130~meV opens between the valence band (VB) at the $\bar{\Gamma}$ point and conduction band (CB) at the $\bar{M}$ point. In the lattice, the normal hexagonal symmetry undergoes a $2a \times 2b \times 2c$ periodic distortion and hence the original BZ is halved in all three crystallographic directions. While a purely electronic hypothesis assumes exciton condensation as the key underlying mechanism~\cite{monney2011exciton,rohwer2011collapse}, other studies emphasized the important role of the lattice in CDW fluctuations~\cite{rossnagel2002charge}. The cooperation of both, excitons and phonons, to develop the CDW phase is a more recent alternative scenario ~\cite{porer2014non,van2010exciton,hedayat2019excitonic,knowles2020fermi,burian2020structurally}. In the following, we show that combining TR-ARPES and TRR provides a unique insight into the CDW formation process in 1$\it{T}$-TiSe$_{2}$.

\subsection{The results of TRR experiment}\label{secTRR}

Figure~\ref{Fig2}a shows the temperature dependent oscillatory response of TRR below the CDW phase transition temperature, $\textit{T}_{\textrm{CDW}}=202$~K. We removed the incoherent electronic part (first term in Eq~\ref{trr2}) by subtracting an exponential function from TRR measurements. We calculate the photo-generated carrier density through $n_{e}=F(1-R)\alpha/ h\nu$, where $F$, $R$ and $\alpha$ denote to pump fluence, the reflectivity and  penetration depth in 1$\it{T}$-TiSe$_{2}$ for the photon energy of the pump~\cite{hedayat2020non}. We estimate $R=0.6$ and $\alpha=15$~nm from Refs.~\cite{reshak2003electronic,velebit2016scattering}. For all curves in Fig~\ref{Fig2}a, the pump-generated carrier density is $n_{e}=2.77\times 10^{18}$~cm$^{-3}$. The TRR measurements at 80~K  display clear oscillations (dashed lines in Fig~\ref{Fig2}a mark the peaks). For the higher temperatures, the oscillations have a smaller amplitude and exhibit stronger damping. Furthermore, a closer inspection reveals that the modulation period increases with temperature (see the difference between dashed lines and the peak marks on the 110~K and 150~K curves). Accordingly, Figs~\ref{Fig2}b-d summarize the frequency and amplitude variations as a function of temperature. The frequency of all oscillations is about $3.2$~THz  (see Fig~\ref{Fig2}b) close to the  frequency of $\textit{A}^{*}_{\textrm{1g}}$ phonon (CDW) mode~\cite{snow2003quantum}. Figure~\ref{Fig2}c and ~\ref{Fig2}d demonstrate that the frequency and amplitude of TRR oscillations decrease with increasing temperature. The trend is similar to a BCS-like temperature dependence of the energy gap~\cite{chen2016hidden,chen2015charge}. We measure TRR with a low photo-induced carrier density to remain close to the equilibrium condition and avoid complete melting of CDW. Our data confirm a strong link between the TRR oscillation and $\textit{A}^{*}_{\textrm{1g}}$ phonon, and subsequently with the structural development of the CDW phase, in line with Ref.~\cite{mohr2011nonthermal}.

We showed the capability of TRR to detect signatures of the PLD by the presence of $\textit{A}^{*}_{\textrm{1g}}$ phonons which are observed only in CDW phase. By increasing the photocarrier density to $6.45\times 10^{18}$~cm$^{-3}$ and exploring a wider range of temperatures, from 80~K to 300~K, we can reveal the interplay between coherent modes in the CDW and normal phases. Two distinct responses in TRR are seen in Figs~\ref{Fig3}a and b, separated by a temperature close to $\textit{T}_{\textrm{CDW}}$. Interestingly, each panel shows only one frequency, a characteristic oscillation which matches the $\textit{A}_{\textrm{1g}}$ or $\textit{A}^{*}_{\textrm{1g}}$ phonons of the normal lattice and the superlattice, respectively~\cite{holy1977raman}. The corresponding Fourier transforms are shown in
Fig~\ref{Fig3}c and Fig~\ref{Fig3}d. When the lattice temperature increases and the sample remains close to the equilibrium state (low photoexcitation regime), the dominant mode in the TRR oscillations switches across $\textit{T}_{\textrm{CDW}}$.
\begin{figure*}
\centering
\includegraphics[scale=1.15]{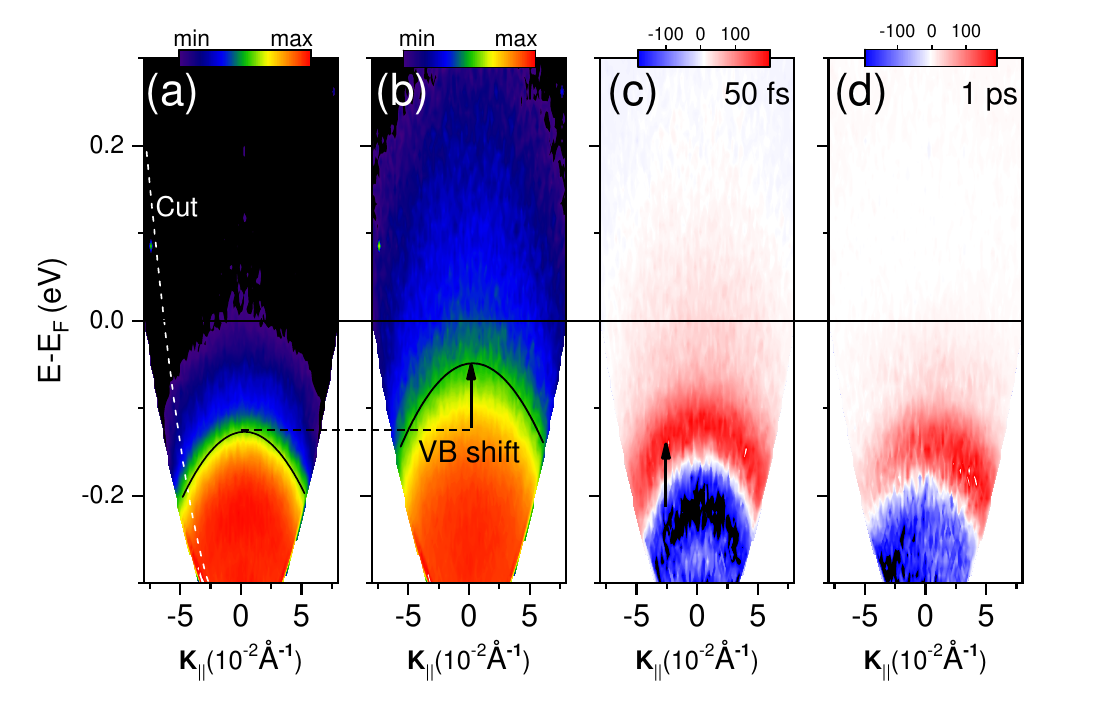}
\caption{TR-ARPES maps of 1$\it{T}$-TiSe$_{2}$ before (a) and at 50~fs after (b) pump illumination in HER. The horizontal line indicates the Fermi level, $\it{E}_{F}$. In (a), the white dashed line indicates a constant angle cut through the spectra where the VB dynamics were extracted. In panel (b) the arrow shows the VB shift towards higher energies. (c) and (d) are differential TR-ARPES maps showing the photoinduced changes at 50~fs and 1~ps delays, created by subtracting a spectrum at negative pump-probe delay. The red and blue colors indicate an increase or decrease of spectral weight, respectively.}
\label{Fig5}
\end{figure*}
The TRR experiment performed close to equilibrium shows an obvious correspondence between the PLD evolution (through detection of different phonons) and temperature. Previous studies showed that the out-of-equilibrium interactions are intricate and need to be comprehended differently from the equilibrium condition in correlated electron systems~\cite{giannetti2016ultrafast}. For example, in 1$\it{T}$-TiSe$_{2}$, the CDWs are melted by a laser pulse with only half of the energy necessary to reach the normal phase thermally~\cite{porer2014non}. Therefore, we performed the TRR measurements at 80~K and increased the pump fluence in order to non-thermally suppress the CDW phase. Figure~\ref{Fig4}a display the oscillatory part of TRR for three photocarrier densities of $1.01\times 10^{19}$~cm$^{-3}$, $7.83\times 10^{19}$~cm$^{-3}$ and $1.21\times 10^{20}$~cm$^{-3}$, referred to respectively as low, medium and high excitation regimes (LER, MER and HER)  at 80~K. For comparison, we also carried out the experiment at 300~K with the photocarrier density of $2.43\times 10^{20}$~cm$^{-3}$. In the LER case, we detect the $\textit{A}^{*}_{\textrm{1g}}$ CDW mode similar to what we extracted in Figs~\ref{Fig2} and ~\ref{Fig3}. To estimate the damping time constant ($\tau_{\mathrm{d}}$) and period of oscillations ($T$), we fit the data with the second term of Eq.~\ref{trr2} after convolution with the instrumental response function of the experiment. Table~\ref{tab1} shows the parameters obtained from the fits. From this, (i) we identified two distinct oscillation frequencies at LER and HER and (ii) we found out that the damping time increases with pump fluence. The frequency change occurs at a carrier density of  $n_{\mathrm{th}}\approx7-8\times 10^{19}$~cm$^{-3}$. Figure~\ref{Fig4}b presents the FTs of TRR oscillations. The corresponding frequencies are $3.24$~THz for LER, $2.95$ and $5.95$~THz for MER and $6.01$~THz for HER. The extracted TRR oscillation frequencies match the $\textit{A}^{*}_{\textrm{1g}}$ (CDW) and $\textit{A}_{\textrm{1g}}$ (normal) phonon modes reported in Ref.~\cite{holy1977raman} showing a shift of the dominant phonon when the photogenerated carrier density exceeds  $n_{\mathrm{th}}$. The redshift seen in $\textit{A}^{*}_{\textrm{1g}}$ when the pump fluence is increased is in good agreement with Ref.~\cite{mohr2011nonthermal}. The observed mode switch from $\textit{A}^{*}_{\textrm{1g}}$ to $\textit{A}_{\textrm{1g}}$ is the signature of melting of the lattice-order associated with the CDW. This evidence raises the question whether, in the HER, the CDW is optically melted. To answer this question we must perform a complementary experiment, because the TRR measurements alone cannot reveal the link between the mode switch and the CDWs. The most direct approach to obtain the CDW magnitude is to probe the energy gap, which is related to the order parameter, using TR-ARPES.
\begin{figure*}
\centering
\includegraphics[scale=1.2]{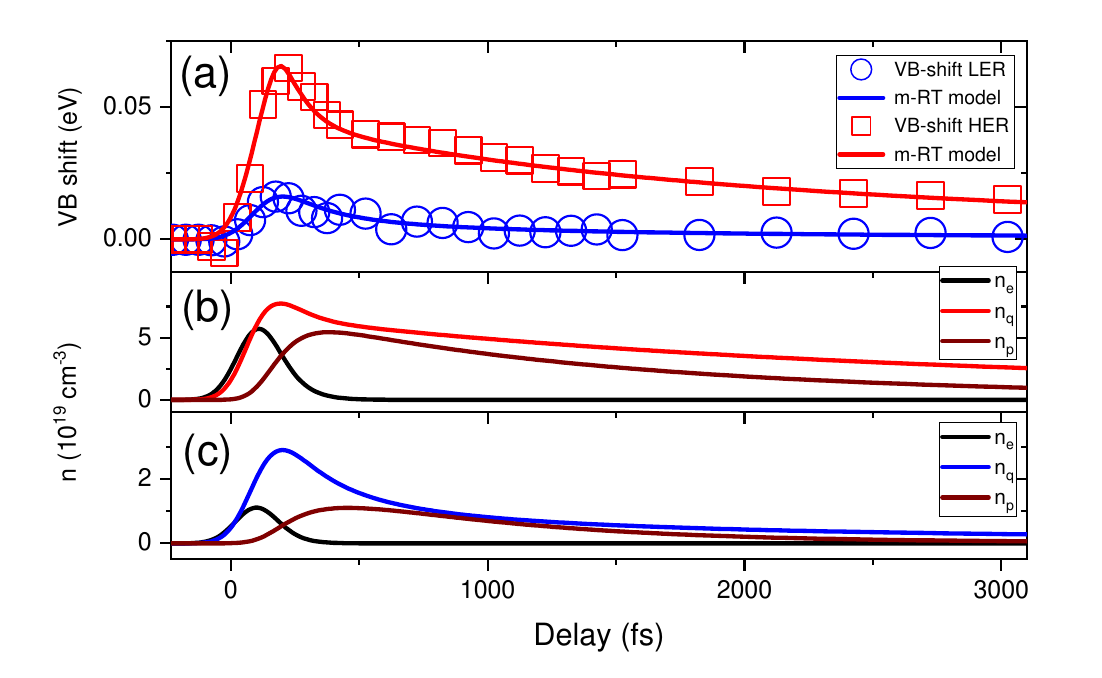}
\caption{(a) VB dynamics as a function of the pump-probe delay for two excitation regimes, $2.86\times 10^{19}$~cm$^{-3}$ (LER) and $1.15\times 10^{20}$~cm$^{-3}$ (HER) at the constant angle cut shown in Fig~\ref{Fig5}a. In the HER, the VB shift is not completely recovered after a few ps. The modified Rothwarf-Taylor (m-RT) model can fit the VB dynamics for both fluences (solid lines). The dynamics of free carriers ($n_{e}$), QPs ($n_{q}$) and SCPs ($n_{p}$) for HER (b) and LER (c) extracted from m-RT model.}
\label{Fig6}
\end{figure*}
\begin{table*}
\caption{\label{tab1} The period ($T_{i}$) and damping time ($\tau_{\mathrm{d,i}}$) of the TRR response in low, medium and high photoexcitation regimes (LER, MER and HER). The parameters were extracted from fits in Fig~\ref{Fig5}.} 
\footnotesize\rm
\begin{tabular*}{\textwidth}{@{}l*{5}{@{\extracolsep{0pt plus12pt}}l}}

& Photocarrier density[cm$^{-3}$]   & $T_{1}$ ($A_{1g}^{*}$)[fs] & $\tau_{\mathrm{d,1}}$[fs] & $T_{2}$ ($A_{1g}$)[fs] & $\tau_{\mathrm{d,2}}$[fs] \\ \hline
 LER  & $1.01\times 10^{19}$ & $315\pm6$ & $396\pm120$ & &  \\ 
MER & $7.83\times 10^{19}$ & $333\pm10$ & $582\pm201$  & $166\pm1$ & $692\pm102$ \\ 
 HER  & $1.21\times 10^{20}$ &  &   & $162\pm2$ & $954\pm182$ \\

\end{tabular*}
\end{table*}

\subsection{TR-ARPES results}
To complement the observations made by TRR, we carried out TR-ARPES measurements with the pump fluence in both the LER and HER. Figure~\ref{Fig5}a and ~\ref{Fig5}b show the TR-ARPES maps exhibiting the VB dispersion at $\bar{\Gamma}$  before and after photoexcitation. Due to the low photon energy of our probe (6 eV), we can only detect a portion of the BZ surrounding the $\bar{\Gamma}$ point, therefore, we cannot directly observe the changes in the energy gap. Nevertheless, since the opening of the CDW gap is \textit{mainly} due to the shift of the VB towards larger binding energies ~\cite{chen2016hidden}, we attribute the VB shift at the $\bar{\Gamma}$ point (see arrow in Fig~\ref{Fig5}b) to the gap closure process. Therefore, we can investigate the gap dynamics at the $\bar{\Gamma}$ point by tracking the VB binding energy as a function of delay. Figure~\ref{Fig5}c and~\ref{Fig5}d display the pump-induced variations in TR-ARPES spectra at 50~fs and 1~ps delays. The differential maps are obtained by subtracting the unperturbed spectra from the spectra at given delays. The red (blue) color denotes an increase (decrease) in the spectral weight, revealing a shift or broadening of the VB after photoexcitation which lasts more than 1~ps. In the following, we analyze the dynamics of the VB shift at a constant angle cut as shown in Fig~\ref{Fig5}a. Figure~\ref{Fig6}a displays the VB dynamics for a photocarriers density of $2.86\times 10^{19}$~cm$^{-3}$ (LER) and $1.15\times 10^{20}$~cm$^{-3}$ (HER). In LER, the gap is reduced upon photoexcitation and fully recovers after about 2 ps. In agreement with TRR data in Fig~\ref{Fig4}a, we partially perturb the CDW but its subsequent reestablishment is fast in the presence of a well-established PLD. Therefore, we see the dominant $\textit{A}^{*}_{\textrm{1g}}$ phonon mode clearly in TRR response. However, the complete suppression of CDW order in the high fluence regime as suggested by the TRR results is not confirmed by the TR-ARPES experiment as seen in Fig~\ref{Fig6}a. In the HER, at early delays, the gap is partially closed by the energy shift of the VB towards $\textit{E}_{\textrm{F}}$ (the maximum VB shift of about 65~meV). However, in contrast to LER, the complete gap recovery is not achieved within the first picoseconds after excitation. After 2 ps, the gap reopens (recovery of $\sim40$~meV) and the CDW phase is partially reestablished. Therefore, the TR-ARPES measurements do not support the hypothesis of complete CDW melting at high fluence as could have been inferred from TRR. Yet, the TR-ARPES data contain additional important information. First, the VB shift recovery cannot be described with a single exponential function for both LER and HER. This fact reveals that different channels must be considered to model the gap recovery behavior. Second, at long delays, there is a crucial difference between the two pump fluence regimes. In HER, where the TRR reveals $\textit{A}_{\textrm{1g}}$ normal phonons, an interaction among quasiparticles prevents the complete gap reopening.  We argue that this interaction is related to phonons, since (i) it is accompanied by the switch of the phonon modes (see Fig~\ref{Fig4}a) and (ii) it occurs in the VB dynamics with the characteristic time scale of phonon interactions (see Fig~\ref{Fig6}a).

\subsection{An adapted model to unify TRR and TR-ARPES results}
We developed a model that can explain the microscopic picture of electronic and lattice interactions allowing us to clarify all experimental outcomes. With the model, we fit the dynamics of the order parameter obtained by TR-ARPES measurements. We then explain the remarkable correlation of the output of the model with the phonon analysis of TRR. A model inspired by the work of Rothwarf and Taylor~\cite{rothwarf1967measurement} is used to simulate the gap dynamics presented in Fig~\ref{Fig6}a. In the modified Rothwarf and Taylor (m-RT) model, we assume the gap is governed by electron-hole pairs in the excitonic insulator scenario for the CDW mechanism in 1$\it{T}$-TiSe$_{2}$~\cite{monney2009spontaneous,rohwer2011collapse}. The gap dynamics, $\Delta (t)$, is related to the broken exciton pairs which form QPs via $\Delta (t) = \Delta_{T_{0}}\sqrt{1-n_{q}(t)/n_{c}}$~, where $\Delta_{T_{0}}=130$~meV for the initial temperature in our TR-ARPES experiment $T_{0}=80$~K~\cite{chen2016hidden}, $n_{q}$ is the QP density and $n_{c}=1.15\times 10^{20}$~cm$^{-3}$ is the critical QP density of 1$\it{T}$-TiSe$_{2}$~\cite{hedayat2019excitonic,wilson1978infrared,li2007semimetal}. The model demonstrates the role of phonons in CDW formation through the time-dependent electron, exciton and phonon interactions captured by the following equations.  

\begin{align} 
&\frac{\mathrm{d}{n}_{e} }{\mathrm{d} t}=-g_{ee}n_{e}-g_{ep}n_{e}-g_{th}n_{e}+P(t),\label{eq4}\\
&\frac{\mathrm{d}{n}_{q} }{\mathrm{d} t}=g_{ee}n_{e}+\eta n_{p}-2Rn_{T}n_{q}-Rn_{q}^{2},\label{eq5}\\
&\frac{\mathrm{d}{n}_{p} }{\mathrm{d} t}=g_{ep}n_{e}-\eta n_{p}/2+Rn_{T}n_{q}+Rn_{q}^{2}/2-\gamma n_{p} \label{eq6}.
\end{align}

\begin{figure*} 
\centering
\includegraphics[scale=0.88]{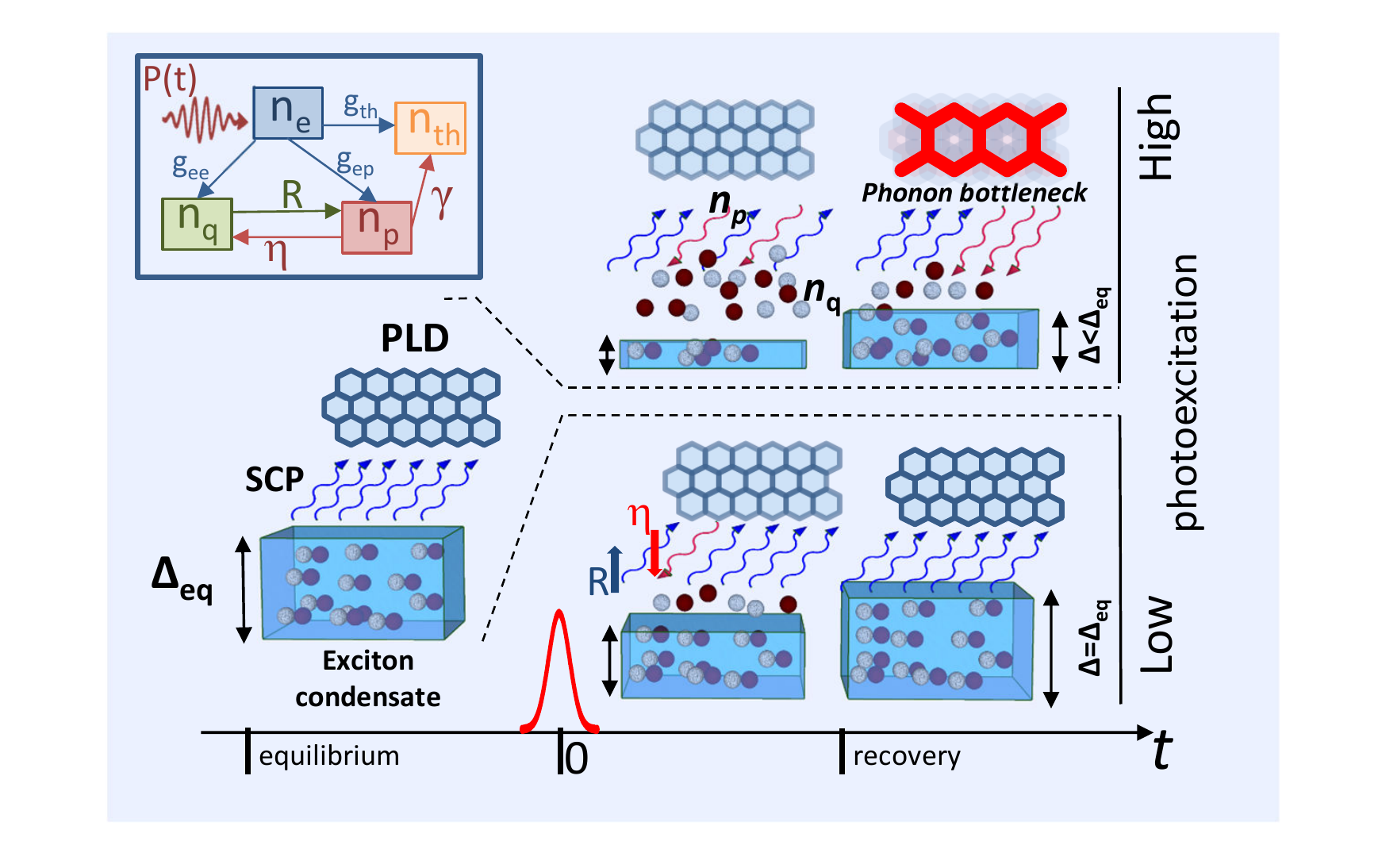}
\caption{A schematic representation of the CDW mechanism governed by exciton and phonon correlations. At equilibrium, before zero delay time, excitons in the condensate (bound electron-hole pairs) and the PLD are responsible for the energy gap in the equilibrium condition, $\mathit{\Delta }_{\textrm{eq}}$. The SCPs (the blue wave arrows) are interacting with both excitons and the PLD. Upon pumping, for both low and high excitation regimes, the SCPs are excited (the red wave arrows). Also, the electron-electron scatterings break exciton pairs and reduce the exciton condensate and consequently the energy gap. This process increases the QP density. QP population (the excitons out of condensate) increases for higher photoexcitation. QPs interact with SCPs with rates R, $\eta$ (red and blue arrows). Excited SCPs (red wave arrows) break the exciton pairs. On the other hand, the SCPs (blue wave arrows) are the main contributors to aid exciton recombination. At higher photoexcitation densities, the SCPs cannot relax due to a strong bottleneck effect and continuously disturb the exciton recombination, and hence the gap is only partially recovered. On the contrary, at low excitations, the SCPs efficiently contribute to CDW recovery, and the gap recovers completely. The inset (top left corner) depicts the interactions between different terms in Eqs.~\ref{eq4}-\ref{eq6}.}
\label{Fig7}
\end{figure*}

Here ${n}_{e}$ and ${n}_{p}$ denote the electron and SPC densities as a function of time, P(t) describes the free carrier generation rate by the pump, $g_{ee}$ and $g_{ep}$ are electron-electron and electron-SCP scattering rates, respectively, and ${n}_{T}$ is the thermal QP density. In the following, we clarify the role of QPs recombination rate ($\mathit{R}$), SCPs relaxation rate through QPs ($\eta$) and anharmonic decay rate of SCPs to the thermal bath ($\gamma$). Figure~\ref{Fig7} (inset, top left hand corner) illustrates all interactions in Eqs.~\ref{eq4}-\ref{eq6}. Our simulations fit the experimental results (Fig~\ref{Fig6}a) and reveal the dynamics of ${n}_{e}$, ${n}_{q}$ and ${n}_{p}$ (Figs~\ref{Fig6}b and ~\ref{Fig6}c) for HER and LER. The results unravel the important role of phonons in the formation of CDWs through their strong interactions with QPs as follows.

Some of the phonons which are preferentially coupled to excitons, i.e. SCPs, are the key ingredient to CDW formation and gap opening. The SCPs and QPs (the excitons out of condensate) strongly interact. After photoexcitation, the population of QPs increases due to the breaking of the exciton pairs by free carrier scattering. Consequently, the gap partially or completely collapses. The QPs can again form excitons by exciting SCPs. In Eqs.~\ref{eq5} and ~\ref{eq6} the rate $\mathit{R}$ determines this process aiding QPs to recombine and join the condensate. Therefore, SCPs assist to the CDW phase reestablishment and the energy gap recovery. On the other hand, excited SCPs can break excitons and increase the QPs populations ($\eta$ in Eqs.~\ref{eq5} and ~\ref{eq6}). This process destabilizes the CDW. The SCPs can relax by anharmonic decay with other phonons~\cite{giannetti2016ultrafast} with the $\gamma$ rate in Eq.~\ref{eq6}. Therefore, $\eta$ and $\gamma$ are two relaxation channels of SCPs. The former creates the QPs from excitons resulting in CDW suppression, the latter reduces the excited SCPs population indirectly promoting CDW recovery. Therefore, for $\eta / \gamma > 1$ the CDW fluctuations are partially quenched by excited SCPs, showing a bottleneck in the gap recovery. From the fit of Fig~\ref{Fig6}a, the bottlenecks for LER and HER are obtained as $\eta / \gamma= 0.15$ and $9.6$, respectively. In the HER, the intense pump pulse promotes large SCPs population by electron-phonon scattering ($g_{ep}$) and QPs recombination ($\mathit{R}$). This process leads to the excited SCPs density ($n_{p}$) that can be absorbed by QPs, resulting in higher $\eta$ rate. On the other hand, the high carrier scattering rate increases the energy of the phonon bath and reduces the PLD intensity, directly by dissipation term ($g_{th}$) or indirectly through the excited SCPs ($\gamma$). Consequently, the decay of SCPs becomes increasingly difficult leading to a decrease of $\gamma$ rate. Both processes give rise to the strong bottleneck effect at HER.

So far, we showed that the m-RT model captures the gap dynamics obtained from TR-ARPES experiments. Remarkably, we find that the results of the simulations are consistent with the TRR findings. The model identifies a significant change in CDWs from LER to HER, owing to a difference in phononic contributions (bottleneck). Similarly, TRR discloses an analogous switch at the crossover of two regimes, $n_{\mathrm{th}}\approx7-8\times 10^{19}$~cm$^{-3}$ (Fig~\ref{Fig4}). In particular, the obtained anharmonic decay-time of SCPs $1/\gamma$ (570~fs and 1070~fs for LER and HER) are closely linked to the progressive damping times of phonons extracted from TRR response (Table~\ref{tab1}, 396~fs to 954~fs).

Fig~\ref{Fig7} summarizes the various interactions captured by the model to illustrate how excitons and phonons cooperatively stabilize the CDW phase. At equilibrium, the energy gap, $\mathit{\Delta }_{\textrm{eq}}$  at 80~K  is governed by the excitons in the condensate, while it is strongly coupled to SCPs and lattice. Upon pumping, electron-electron scattering events produce two main effects; (i) they break some exciton pairs in  the condensate creating the QP population and (ii) they increase the excited SCPs population (the red wave arrows). Notice that $\Delta$ is reduced for both LER and HER. At later times, the gap $\Delta$ recovers for low-intensity pump but remains partially open for higher fluences. As depicted, the rates $\mathit{R}$, $\eta$ and $\gamma$ all contribute to the establishment of CDW order. When the number of excited SCPs increases (red wave arrows), the rate $\eta$ increases (higher probability of pair breaking) and $\gamma$ decreases (lower probability of SCPs decay) leading to a strong bottleneck in the HER. The top right scheme of Fig~\ref{Fig7} shows that these bidirectional interactions disturb the complete gap recovery. 

\section*{Conclusion} 
Our results on 1$\it{T}$-TiSe$_{2}$ show that the combination of ultrafast spectroscopy techniques provides a novel experimental approach for investigating the unexplored complex non-equilibrium interactions in strongly correlated materials. To demonstrate the capability of such combined techniques, we used complementary TRR and TR-ARPES probes to disclose the real-time snapshots of complex phenomena triggered by the same optical excitation. This unique approach allowed us to develop a complete picture of the non-thermal response of separate electron and lattice subsystems, and their interactions. We revealed the mechanisms behind CDW formation in 1$\it{T}$-TiSe$_{2}$ and highlighted the important role of phonons in the recovery following optical excitation. To summarize, we found that the SCPs are strongly coupled to excitons, such that any perturbation of one will destabilize the other, and consequently influence the CDW order. Therefore, the CDW phase in 1$\it{T}$-TiSe$_{2}$ is a consequence of both excitonic and phononic interactions. Given these promising results, we believe it is worthwhile to develop and refine combined ultrafast techniques. Our approach can be employed to investigate the detailed mechanisms driving the phase transitions in a wide range of strongly correlated materials~\cite{shao2018recent,Sayers2020Correlation,gedik2007nonequilibrium}.

 \section*{Acknowledgments} 
 G.C, E.C and H.H acknowledge funding from the European Union Horizon $2020$ Programme under Grant Agreement No.~$881603$ Graphene Core $3$, PRIN $2017 - 2017BZPKSZ 002$. C.J.S. acknowledges funding and support from the EPSRC Centre for Doctoral Training in Condensed Matter Physics (CDT-CMP), Grant No. EP/L$015544/1$. J.v.W. acknowledges support from a VIDI grant financed by the Netherlands Organization for Scientific Research (NWO). E.D.C. acknowledges support from Horizon $2020$ (Grant No. $654148$, Laserlab-Europe). S.R.C. acknowledges support from EPSRC under Grant No.
EP/P$025110/2$.  

\def\bibsection{\section*{\refname}} 
\bibliography{bibliography}

\end{document}